\documentclass[12pt]{iopart}

\usepackage[dvips]{graphicx}
\usepackage[dvips]{color}
\usepackage{subfigure}[]
\usepackage[ansinew]{inputenc}

\begin{document}

\title[Long-range stress-engineered ordering of III-V semiconductor nanostructures]
{Long-range ordering of III-V semiconductor nanostructures by shallowly buried dislocation networks}
\author{J Coelho, G Patriarche, F Glas, G Saint-Girons and I Sagnes}
\address{Laboratoire de Photonique et de Nanostructures, LPN-CNRS/UPR20, Route de Nozay,
91460 Marcoussis, France} \ead{jose.coelho@lpn.cnrs.fr}
\begin{abstract}
We account for lateral orderings of III-V nanostructures resulting
from a GaAs/InAs/InGaAs/GaAs sequence grown on GaAs by
metalorganic vapour phase epitaxy at two different temperatures.
For both samples, the ordering is induced by the stress field of a
periodic dislocation network (DN) shallowly buried and parallel to
the surface. This DN is a grain boundary (GB) that forms, between
a thin GaAs layer (on which growth was performed) and a GaAs
substrate joined together by wafer bonding, in order to
accommodate a tilt and a twist between these two crystals; both
these misorientations are imposed in a controlled manner. This GB
is composed of a one-dimensional network of mixed dislocations and
of a one dimensional network of screw dislocations. For both
sample, the nanostructures observed by transmission electron
microscopy (TEM) and atomic force microscopy are ordered by the
underlying DN observed by TEM since they have same dimensions and
orientations as the cells of the DN.
\end{abstract}
 \submitto{Journal of
Physics: Condensed Matter}

\section{Introduction}

The control on a wide surface of the size, position and density of
self-assembled quantum dots (QDs) is a requirement for the
increase of the performances of numerous optoelectronic devices,
such as semiconductor lasers, or for the realization of new
devices, such as a single photon source for quantum information
based on an isolated QD. One way to achieve this control is to use
the periodic stress field induced at the surface of a specimen by
a periodic dislocation network (DN) shallowly buried and parallel
to the surface, which generates preferential nucleation sites for
the QDs \cite{Bourret99,Romanov99}. So, by choosing an adequate
periodicity, it should be possible to organize laterally QDs with
identical sizes. Here we study two samples for each of which a
shallowly buried DN accommodates crystallographic misorientations
imposed in a controlled manner between two GaAs crystals (a
substrate and a thin layer) joined by wafer bonding. This method
has the outstanding property of no leading to the formation of any
threading dislocation which could affect the optical properties of
the subsequently grown layers \cite{Liau90}. Moreover, since the
charge carriers are strongly confined in QDs, the optical
sensitivity of the latter to the presence of defects (such as the
dislocations of an underlying DN) should be weak \cite{Gerard96}.
Till now, lateral ordering of nanostructures mediated by an
underlying DN has been reported for metals \cite{Brune98} and for
Germanium on Silicon \cite{Leroy02} but not for III-V materials.
Here, we report a major step towards the long-range lateral
organization of III-V self-assembled QDs, namely long-range
lateral organizations of III-V nanostructures induced by the
buried DNs previously reported. Growth was performed at different
temperatures on two samples, for which, the periodicities of the
DNs are also different.

\section{Experimental procedure}

At the interface between two crystals, a grain boundary (GB)
forms. This GB is constituted of a periodic DN that accommodates
the crystalline discontinuity. Thus, to obtain a shallowly buried
DN, we transfer by wafer bonding a thin crystalline GaAs layer on
a host GaAs substrate, between which we impose controlled
misalignments of their reticular planes (to be detailed below). We
called the resulting structure a `composite substrate'.

First the thin layer (approximately 20 nm thick) is grown by
metalorganic vapour phase epitaxy (MOVPE) on a sacrificial GaAs
substrate. Two Al$_{0.9}$Ga$_{0.1}$As etch-stop layers separated
by a GaAs buffer layer are also grown between the thin GaAs layer
and the sacrificial substrate in order to allow the removal of the
latter (as well as the AlGaAs layers) by wet selective chemical
etching after the bonding. We use two etch-stop layers rather than
only one to better control the removal of the sacrificial
substrate and obtain a surface as smooth as possible.

After cleaning and deoxidizing, this structure and the host
substrate are superposed with controlled crystalline misalignments
imposed between them. These misalignments are a twist (i.e. a
rotation around an axis orthogonal to the interface) and a tilt
(i.e. a rotation around an axis lying in the interface). The tilt
is established by using commercial vicinal wafers: their surfaces
are disoriented by $0.3 \pm 0.1^{\circ}$, around an in plane
$<100>$ direction, with respect to the (001) plane. On the other
hand, to control the twist, we first cut with a saw square pieces
of a wafer to obtain sides having the desired disorientation with
respect to the $<$110$>$ cleavage directions. We then put in
contact and align the sides of a sawn square and of a square
simply cleaved along the $<$110$>$ directions -- these two squares
are the crystals that will be bonded. This method allows twist
control to within $\pm 0.1^{\circ}$. Considered independently, a
twist between two crystals is accommodated by a square
two-dimensional (2D) network of screw dislocations, while a tilt
is accommodated by a one-dimensional (1D) network of mixed
dislocations oriented along the tilt axis (the line orthogonal to
the maximum slope of the interface). The two crystals were
superposed so that the maximum slope of their surfaces are
orthogonal and thus the resulting maximum slope of the interface
is along a $<110>$ direction. For these two DNs, the periodicity
$D$ is:
\begin{equation}\label{eq:eq1}
D=\frac{b^\prime}{2\sin{\theta/2}}\;,
\end{equation}
where $b^\prime$ is the modulus of the Burgers' vector component
allowing the accommodation of the misorientation considered and
$\theta$ is the misorientation angle. In III-V materials, for
screw dislocations, $b^\prime$ is the entire modulus of the
Burgers vector, equal to $\frac{a}{\sqrt{2}}$ (where $a$ is the
lattice parameter of the material), while for mixed dislocations
it is the Burger's vector component normal to the boundary plane:
$b^\prime=\frac{a}{2}$.

Afterwards, this stack is annealed during one hour, at
600$^{\circ}$C, under nitrogen flow. During this heating, a
mechanical pressure (between 10 and 100 kg/cm$^2$) is applied on
the stack in order to maintain the surfaces of the two crystals in
contact. Indeed, contrary to Si/Si bonding for example, deoxidized
and flat GaAs surfaces do not bond at room temperature when they
are simply put in contact. On the other hand, the difficulty to
impose a homogeneous pressure on a wide surface limits the
dimensions of our composite substrates. However, we succeeded to
increase their surface area from 1 cm$^2$ to 4 cm$^2$. During the
annealing, covalent bonds form at the bonding interface, while a
DN forms to accommodate the misalignments.

\begin{figure}[!htb]
\centering
\includegraphics*[10mm,210.1mm][85.3mm,287mm]{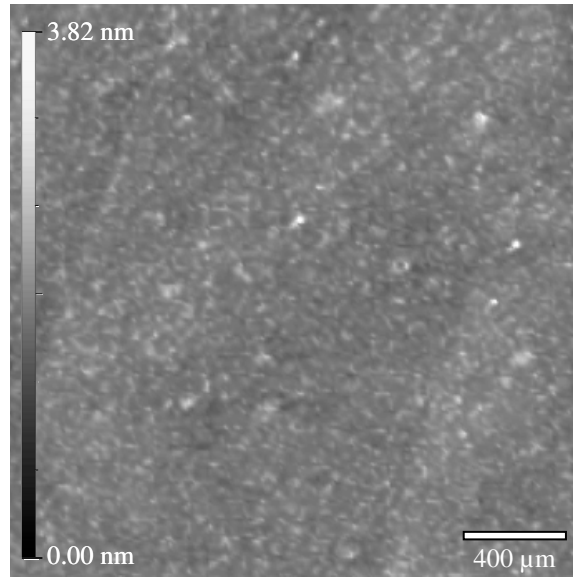}
\caption{AFM image of a composite substrate before growth.}
\label{fig:Fig1}
\end{figure}

After, the sacrificial GaAs substrate is mechanically thinned and
then etched by a stirred citric acid solution obtained by
dissolving 50 g of citric acid in 50 cm$^3$ of deionized water and
adding 20 cm$^3$ of 30\% H$_2$O$_2$. On the other hand, the AlGaAs
layers are eliminated with a 5\% hydrofluoric solution, to leave
only the thin GaAs layer bonded to the host substrate. Figure
\ref{fig:Fig1} illustrates the typical resulting surface. It is
very flat: the root mean square (RMS) roughness of 0.28 nm is
similar to that of regular GaAs substrates. Notice that for the
chemical etchings to be selective it is important that the etched
materials be undoped.

Finally, on two such composite substrates, after a 9 min annealing
at 650$^{\circ}$C which purpose was to evaporate the surface
oxide, we deposited by MOVPE a III-V multilayer that gives rise on
regular substrate to the formation of QDs \cite{Saint-Girons02},
in order to observe the effect, on this growth, of the strain
field of the buried DN. The growth sequence was
GaAs/InAs/In$_{0.15}$Ga$_{0.85}$As/GaAs. It was performed at 470
$^{\circ}$C (sample A) or 450$^{\circ}$C (sample B).

\section{Results and discussion}

\begin{figure}[!htb]
\centering
\includegraphics*[10mm,221.4mm][85mm,287mm]{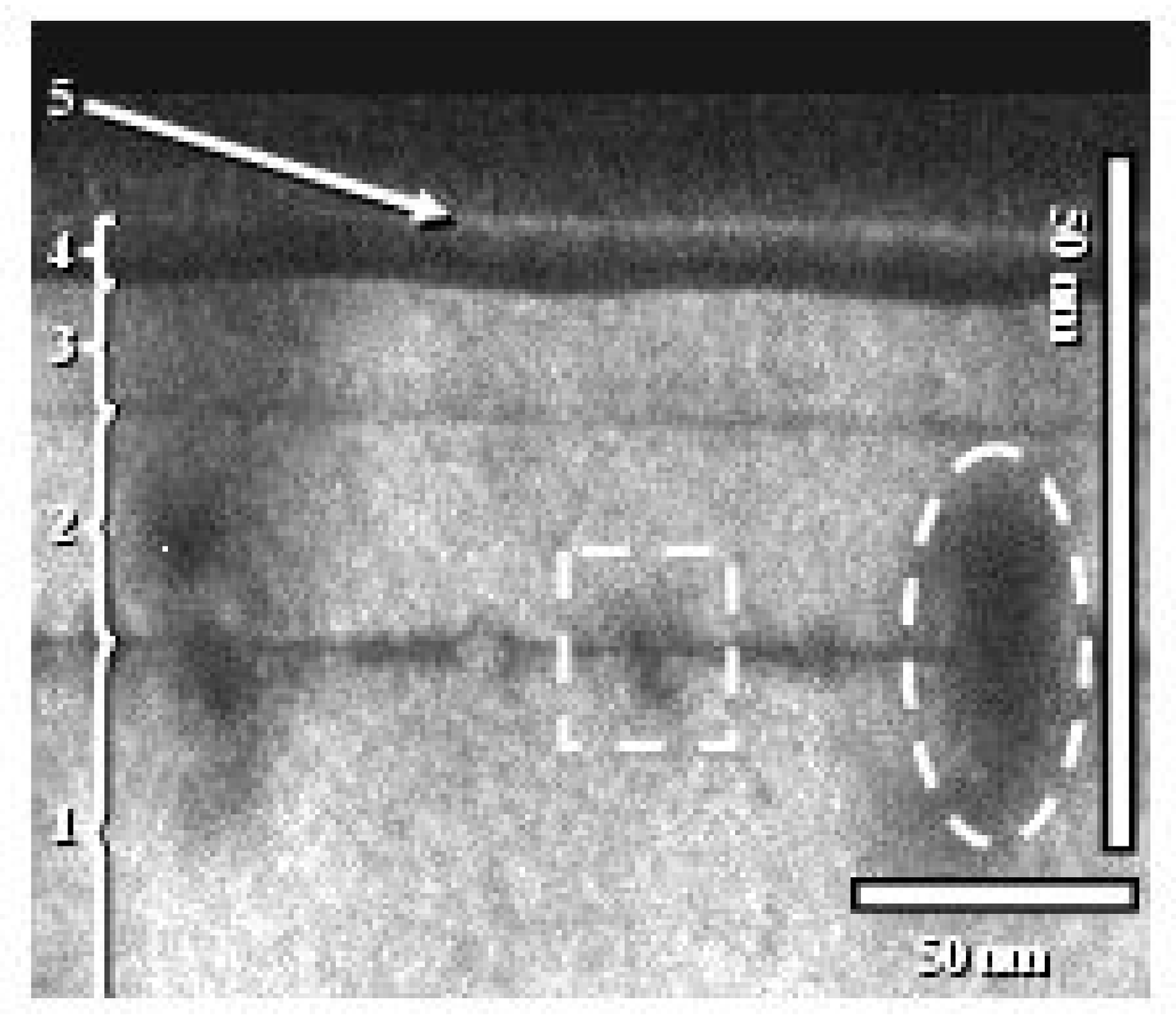}
\caption{TEM 002 dark-field cross-sectional image of sample A. The
different layers are detailed in the text. Oval and rectangle
indicate respectively an interface dislocation and an interface
cavity. Note the different horizontal and vertical scales chosen
to enhance the undulations of layers \#3 and 4.} \label{fig:Fig2}
\end{figure}

Cross-section transmission electronic microscope (TEM)
observations of the structures showed that they are different from
those obtained on regular substrates (the phenomenon is
illustrated in figure \ref{fig:Fig2} for sample A).
Starting from the bottom, we find the host GaAs substrate (\#1)
and the bonded GaAs layer (\#2). Their interface is the GB, where
the large dark spots are due to the strain fields around the
dislocations and the small ones to cavities (resulting from the
non planarity of the surfaces put in contact) or to segregated
impurities. As expected, the dislocations remain confined to the
GB and do not propagate in the surrounding layers. The grown
layers are above layer \#2. No QD is observed in this sample.
Nevertheless, both the GaAs buffer layer (\#3) and the InGaAs
alloy layer (\#4; which results from the intermixing between the
InAs and In$_{0.15}$Ga$_{0.85}$As deposited layers) exhibit
thickness modulations, to be discussed below. Finally, a thin GaAs
layer (\#5) covers the entire structure (the weak contrast above
the latter is due to glue). The presence of a dark line at the
\#2/\#3 GaAs/GaAs homointerface might seem surprising. However,
the top of layer \#2, on which growth is started, is obtained by
chemical etching and cannot have the quality of standard
`epi-ready' wafers. Moreover, secondary ion mass spectroscopy
shows that impurity levels as low as 10$^{18}$ cm$^{-3}$ suffice
to produce such features. Finally, from the image intensity ratio
between the In$_x$Ga$_{1-x}$As and the GaAs in figure
\ref{fig:Fig2}, and using our previous work \cite{Patriarche04},
we determined the average indium composition of layer \#4:
x=0.31$\pm$0.02.

From such images, it appears readily that the thickness
modulations, which affect both the GaAs buffer and the InGaAs
layer, are not randomly distributed: for instance, thicker InGaAs
grows in the valleys of the GaAs layer. Since moreover their
dimensions, modulation periods and modulation amplitudes are of
the order of between 1 and 100 nm, these features truly constitute
III-V semiconductor nanostructures. These nanostructures are
clearly the direct effect of the underlying dislocations during
growth, and are not mediated by a possible undulation of the
initial growth surface, namely the top of layer \#2; indeed, as
was shown in figure \ref{fig:Fig1}, the latter exhibits a
negligible non-ordered corrugation. However, cross-sections such
as figure \ref{fig:Fig2} allow a detailed study neither of their
organization nor of the relationship between the underlying
dislocations and the nanostructures.

\begin{figure}[!htb]
\subfigure{
\label{fig:Fig3:subfig:a} %
\begin{minipage}[b]{0.5\textwidth}
\centering
\includegraphics*[10mm,238.1mm][75.2mm,287mm]{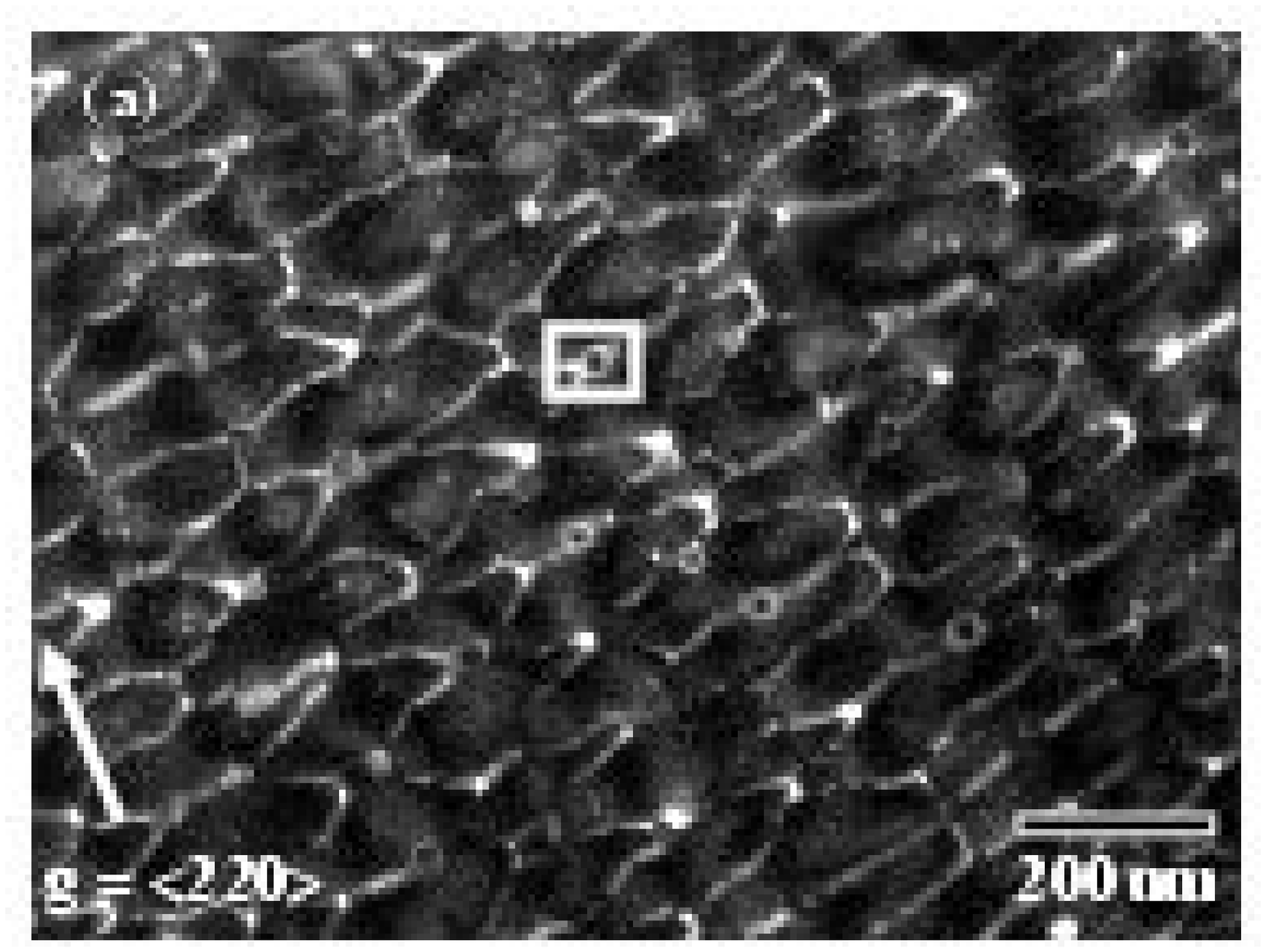}
\end{minipage}}
\subfigure{
\label{fig:Fig3:subfig:b} %
\begin{minipage}[b]{0.5\textwidth}
\centering
\includegraphics*[10mm,
238.5mm][73.3mm,287mm]{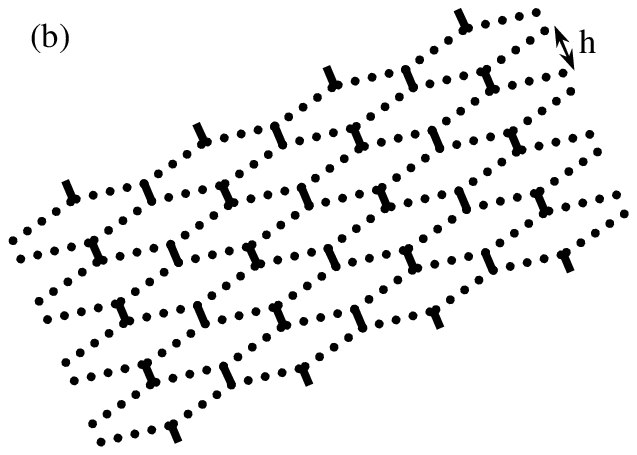}
\end{minipage}}
\caption{(a) TEM dark-field plan-view image of sample A in $<220>$
weak beam condition. An interface cavity is marked by a square.
(b) Schematics of (a) with mixed (dotted lines) and screw (full
lines) dislocations.}
\label{fig:Fig3:subfig} %
\end{figure}

Figure \ref{fig:Fig3:subfig:a} is a TEM plan-view of sample A,
obtained with diffraction vector \textbf{g} along a $<220>$
direction. Such imaging condition reveals chiefly the high strain
field localized close to the dislocation cores for which
\textbf{g.b} is not zero and not the more diffuse strain field
associated with the thickness modulations of the GaAs and InGaAs
layers. The interface cavities are also imaged; though many of
them seem to deviate and pin the dislocations, their distribution
is random and homogenous (in particular, it is not related to the
periodicities of the dislocations). On the other hand, weak beam
images formed with the orthogonal $<220>$ reflection seem to show
only the same dislocations. TEM plan-views taken with the same
imaging conditions on sample B reveal a similar grain boundary,
with slightly different periodicities (to be detailed below).

From detailed observations on similar DNs samples (to be reported
elsewhere), we could identify the DNs of the present GBs. They are
constituted of a 1D network of screw dislocations shifted,
perpendicularly to their line direction, by approximately half a
period each time it crosses a dislocation of the orthogonal 1D
network of mixed dislocations (a schematic representation of
figure \ref{fig:Fig3:subfig:a} is shown in figure
\ref{fig:Fig3:subfig:b}). For sample A, the period of the screw DN
is $261\pm61$ nm, corresponding to a twist of
$0.09\pm0.02^{\circ}$, and that of the mixed DN is $50\pm15$ nm,
corresponding to a tilt of $0.36\pm0.11^{\circ}$. For sample B,
the period of the screw DN is $128\pm20$ nm, corresponding to a
twist of $0.18\pm0.03^{\circ}$, and that of the mixed DN is
$49\pm3$ nm, corresponding to a tilt of $0.33\pm0.02^{\circ}$. All
these experimental values are in agreement with the expected ones.
The shifts of the screw dislocations, as well as the saw-tooth
aspect of the mixed dislocations, arise from energy minimizing
interactions \cite{Hirth82,Benamara95}. The resulting dislocations
cells are roughly hexagonal. The long dimension of these cells is
exactly the period of the screw DN and the short dimension is $h$
(figure \ref{fig:Fig3:subfig:b}): $h=88\pm32$ for sample A and
$h=63\pm11$ nm for sample B.

So, whereas we expected for both samples a 2D network of screw
dislocations allowing to accommodate the twist, here we only
observe a 1D DN. From the detailed studies of our other DNs that
allowed us to identify the dislocations of the present samples, we
could prove that when the mixed dislocations are approximately
oriented along a $<$110$>$ direction (like the screw
dislocations), they accommodate part of the twist by means of
their Burgers vector screw components (that will be detailed
elsewhere). For both samples, we observed such mixed dislocations.
Moreover, since for both cases the mixed dislocations are more
than two times more numerous than the screw dislocations and
though they are two times less efficient than the latter to
accommodate a twist (their Burgers vector screw components are two
times lower than those of the latter), it is possible that they
`replace' the missing 1D screw DN and accommodate the entire
remaining twist. With respect to interface energy, it seems highly
favorable to eliminate totally one half of the standard 2D screw
DN.

\begin{figure}[!htb]
\subfigure{
\label{fig:Fig4:subfig:a} %
\begin{minipage}[b]{0.5\textwidth}
\centering
\includegraphics*[11mm,
233mm][80mm,287mm]{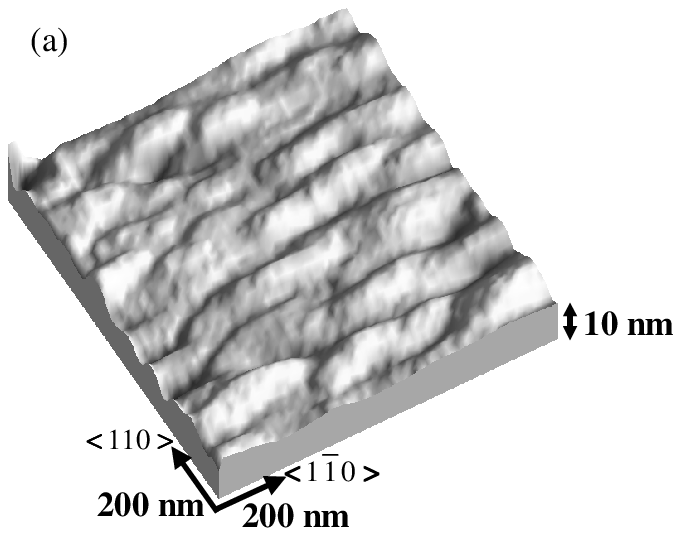}
\end{minipage}}
\subfigure{
\label{fig:Fig4:subfig:b} %
\begin{minipage}[b]{0.5\textwidth}
\centering
\includegraphics*[10mm,
229.9mm][76.5mm,287mm]{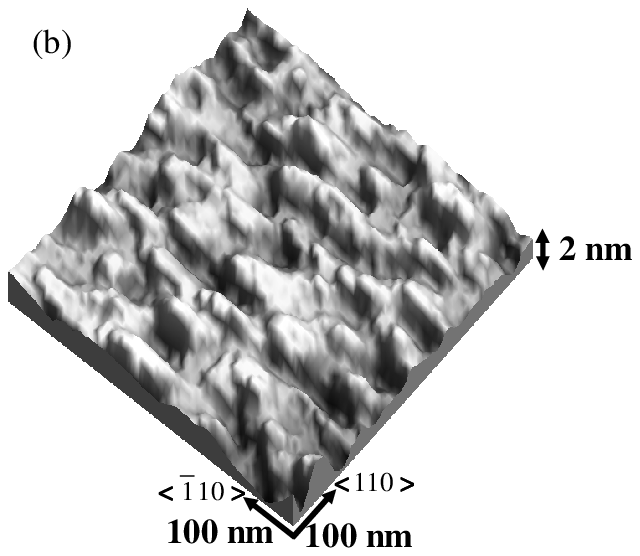}
\end{minipage}}
\caption{AFM images showing the surface corrugation induced by the
organized nanostructures (a) for sample A and (b) sample B.}
\label{fig:Fig4:subfig} %
\end{figure}

Whereas AFM images of composite substrates before growth reveal
only a flat non-ordered surface (figure \ref{fig:Fig1}), AFM
observations on samples A and B after growth (figure
\ref{fig:Fig4:subfig}) reveal the presence of nanostructures.
These nanostructures result from the superposition of the GaAs and
InGaAs nanostructures observed in the cross-section. However since
the GaAs nanostructures are higher than the InGaAs ones, the
nanostructures observed in figure \ref{fig:Fig4:subfig} are
certainly mostly due to the former. These surface nanostructures
are elongated along a $<$110$>$ direction like the cells of the
underlying DNs. Moreover, for both samples, their lateral
dimensions, measured by height profiles taken along the $<$110$>$
directions, are identical to those of the DN cells (to within
experimental uncertainties). So, for both samples, though the
periodicities of their DNs and the growth temperatures are
different, the nanostructures are ordered by these shallowly
buried DNs.

By comparing the two images of figure \ref{fig:Fig4:subfig} and
though they have different scales (adapted to the different
dimensions of the nanostructures), we notice that the surface
nanostructures of sample A seem flatter than those of sample B. We
expected such a difference; indeed, for the latter sample, the
growth temperature was lowered by 20$^{\circ}$C in order to
diminish the diffusion length of atoms and favour the formation of
less flat nanostructures, our aim being to approach the aspect of
conventional QDs. However, the mean height of these nanostructures
is lower than for sample A: 1 nm, against 2 nm for the latter.
This lower value for sample B is not surprising and results from
the lower period of the DN: when dislocations approach each other,
i.e. when the DN period is reduced, the stress induced at the
surface is reduced because of a screening effect \cite{Bourret99}.
In order to determine with certainty which of the samples really
has flatter surface nanostructures, we assimilated the
nanostructures bases to ellipses and calculated the ratios between
their heights and their bases areas. For sample A, the average of
these ratios is $1.6\times10^{-4}$ nm$^{-1}$ while for sample B it
is $2.6\times10^{-4}$ nm$^{-1}$, that is 1.6 times higher than for
sample A. So the surface nanostructures of sample B are indeed
less flat than those of sample A.

\section{Conclusion}

We managed to order III-V nanostructures with the stress field of
periodic shallowly buried DNs. These nanostructures consist of
modulations of GaAs and InGaAs layers due to the stress field of
the dislocations. Moreover, we demonstrated that by modifying the
growth conditions (namely, by reducing the growth temperature), we
can obtain less flat nanostructures. It is a very encouraging
result in the perspective of ordering QDs for applications to
optical emitters. Such ordered QDs should be formed by further
reducing the size of the DN cells, which should lead to the
reduction of the lateral dimensions of the nanostructures.

\ack{We thank C. Mériadec for her expert assistance during the
wafer bonding experiments and C. David for the AFM images. This
work was supported by the Ile de France region, SESAME project \#
1377 and the Conseil Général de l'Essone.}

\end{document}